\documentstyle[preprint,aps,prd]{revtex}
\draft
\title{Comment on entropy bounds and the generalized second law}
\author{M.A. Pelath and Robert M. Wald}
\address{Enrico Fermi Institute and Department of Physics\\
University of Chicago\\
5640 S. Ellis Avenue\\
Chicago, Illinois 60637-1433}
\date{\today}
\begin{document}
\maketitle
\begin{abstract}
In a gedanken experiment in which a box initially containing energy
$E$ and entropy $S$ is lowered toward a black hole and then dropped
in, it was shown by Unruh and Wald that the generalized second law of
black hole thermodynamics holds, without the need to assume any bounds
on $S$ other than the bound that arises from the fact that entropy at
a given energy and volume is bounded by that of unconstrained thermal
matter. The original analysis by Unruh and Wald made the approximation
that the box was ``thin'', but they later generalized their analysis
to thick boxes (in the context of a slightly different
process). Nevertheless, Bekenstein has argued that, for a certain
class of thick boxes, the buoyancy force of the ``thermal atmosphere''
of the black hole is negligible, and that his previously postulated
bound on $S/E$ is necessary for the validity of the generalized second
law. In arguing for these conclusions, Bekenstein made some
assumptions about the nature of unconstrained thermal matter and the
location of the ``floating point'' of the box. We show here that under
these assumptions, Bekenstein's bound on $S/E$ follows
automatically from the fact that $S$ is bounded by the entropy of
unconstrained thermal matter. Thus, a box of matter which violates
Bekenstein's bound would violate the assumptions made in his analysis,
rather than violate the generalized second law. Indeed, we prove here
that no universal entropy bound need be hypothesized in order to
ensure the validity of the generalized second law in this process.

\end{abstract}


\section{Introduction}

A cornerstone of black hole thermodynamics is the {\it generalized
second law} (GSL), which asserts that in any process, the {\it
generalized entropy}
\begin{equation}
S' = S + S_{\rm bh}
\end{equation}
never decreases, where $S$ denotes the entropy of matter outside of
black holes and $S_{\rm bh} = {\cal A}/4$, where ${\cal A}$ denotes
the total surface area of the black hole horizons. (Here and
throughout this paper we use units where $G = c = \hbar = k = 1$.) The
validity of the GSL is essential for the consistency of black hole
thermodynamics and for the interpretation of ${\cal A}/4$ as representing the
physical entropy of a black hole.

It was already recognized at the time the GSL was first
postulated that a potential difficulty arises when one lowers
a box initially containing energy $E$ and entropy $S$ toward a black
hole \cite{bek73}. Classically, a violation of the GSL can be achieved
if one lowers the box sufficiently close to the horizon. A resolution
of this difficulty was proposed by Bekenstein by postulating that the
entropy to energy ratio of any matter put into a box must be subject
to the universal bound \cite{bek81}
\begin{equation}
S/E \leq 2 \pi R
\label{S/E}
\end{equation}
where $E$ denotes the energy in the box, and $R$ denotes some suitable
measure of the size of the box. Naively, at least, such a bound would
rescue the GSL by preventing one from lowering a box close enough to a
black hole to violate it.

However, an alternative resolution of the apparent difficulty with the
GSL was given in \cite{uw82}. There it was noted that there is a
quantum ``thermal atmosphere'' surrounding a black hole, which
produces a large ``buoyancy force'' on a box when it is slowly lowered
very close to the horizon. When this buoyancy force is taken into
account, the optimal place to drop such a box into a black hole no
longer is at the horizon of the black hole but rather at the
``floating point'' of the box, which lies at a finite distance from
the horizon. When the effects of the buoyancy force on energy balance
are properly taken into account, it was found that the GSL always
holds in this process \cite{uw82}.

The analysis of \cite{uw82} assumed, for simplicity, that the box was
``thin'' in the sense that its proper height, $b$, is small compared
with the scale of variation of the redshift factor, $\chi$, i.e., $b
\ll \chi (d \chi/ dl)^{-1}$, where $l$ denotes proper distance from the
horizon. This analysis was then generalized to the case of ``thick
boxes'' in \cite{uw83}, although this generalization was done in the
context of a slightly different process, wherein, rather than dropping
the box into the black hole, the contents of the box are allowed to
``leak out'' as the box is raised. Nevertheless, several years ago
Bekenstein argued \cite{bek94} that for boxes with $b$ at least as
large as $A^{1/2}$ (where $A$ denotes the horizontal cross-sectional
area of the box), the buoyancy effects of the thermal atmosphere are
negligible. He then showed that the bound (\ref{S/E}) must hold for
such boxes in order that the GSL be valid, in apparent contradiction
with the conclusions of the analysis of \cite{uw82,uw83}.

The purpose of this paper is to resolve this apparent
contradiction. In the course of his analysis, Bekenstein \cite{bek94}
made some assumptions concerning the nature of unconstrained thermal
matter and the location of the floating point of the box. Under these
assumptions, it is indeed necessary for the validity of the GSL that
the bound (\ref{S/E}) hold, as Bekenstein found. However, we shall
show that eq.\ (\ref{S/E}) holds {\it automatically} as a consequence
of the same assumptions used to show that it is necessary for the
validity of the GSL. In other words, if one had matter which violated
eq.\ (\ref{S/E}), then Bekenstein's assumption about the nature of
unconstrained thermal matter and/or his assumption about the location
of the floating point of the box could not be correct. 

In the next section, we show that---whether or not eq.\ (\ref{S/E}) is
satisfied---the GSL holds in any process where a (possibly ``thick'')
box, initially containing energy $E$ and entropy $S$, is lowered
toward a black hole and then dropped in. Bekenstein's arguments are
then analyzed in Section 3.

\section{Validity of the GSL for ``thick'' boxes}

It was shown in \cite{uw82} that the bound (\ref{S/E}) is not needed
for the validity of the GSL for the case of a ``thin'' box. The
analysis of \cite{uw82} was generalized to ``thick'' boxes in the
Appendix of \cite{uw83}. However, in \cite{uw83} a slightly different
process was considered (in response to criticisms of \cite{bek83}), in
which the contents of the box are allowed to slowly leak out as the
box is raised. Consequently, the formulas of \cite{uw83} are not
immediately applicable to the present situation where the box is
dropped into the black hole. Thus, in this section, we shall extend
the analysis and arguments given in the Appendix of \cite{uw83} to the
present case.

To begin, in a given region of space outside of the black hole, {\it
unconstrained thermal matter} is {\it defined} to be the state of
matter that maximizes entropy at a fixed volume and energy (as
measured at infinity).\footnote{By contrast, the terminology ``thermal
matter'' would be used to denote matter which is in thermal equilibrium
but which may have additional ``constraints'' resulting, e.g., from
the presence of box walls (which may exclude some modes of excitation
of the matter) or restrictions on the species of particles that are
present.} It should be noted that the properties of unconstrained
thermal matter may depend upon location, i.e., for unconstrained
thermal matter the functional dependence of the entropy density, $s$,
on energy density, $e$, may vary with position outside of the black
hole. We make two assumptions about unconstrained thermal matter: (i)
We assume that unconstrained thermal matter is (locally) homogeneous,
so that the integrated Gibbs-Duhem relation holds \cite{uw82}
\begin{equation} 
e + P - Ts = 0 
\label{gfe} 
\end{equation}
where $T$ is the temperature of the unconstrained thermal matter, and
$P$ is its pressure. (ii) We assume that the ``thermal atmosphere'' of
a black hole is described by unconstrained thermal matter, with
locally measured temperature given by $T = T_{\rm{bh}}/\chi$, where $T
= T_{\rm{bh}} = \kappa/2 \pi$ is the Hawking temperature of the black
hole. Both of these assumptions were also made in Bekenstein's
analysis \cite{bek94}.

Following \cite{uw83} and \cite{bek94}, we now compute the change in
generalized entropy occurring when matter in a thick box is slowly
lowered toward a black hole and then dropped in.  Consider a box of
cross-sectional area $A$ and height $b$, containing energy density
$\rho$ and total entropy $S$. As the box is lowered toward the black
hole, the energy density will depend on both the proper distance, $l$,
of the center of the box from the horizon and the proper height, $y$,
above the center of the box. Following \cite{bek94}, we adopt the
abbreviation
\begin{equation}
\int f(y) dV \equiv A \int_{-b/2}^{b/2} f(y) dy.
\end{equation}

The energy of the box as measured at infinity is
\begin{equation}
E_{\infty}(l) = \int \rho(l,y) \chi(l+y) dV
\label{Einfty}
\end{equation}
where $\chi$ is the redshift factor. The weight of the box at infinity
is
\begin{equation}
w(l) = \int \rho(l,y) \frac{\partial \chi(l+y)}{\partial l} dV.
\end{equation}
The condition that no extra energy is fed into or taken out of the box
as it is lowered is\footnote{If the box is filled with matter in
thermal equilibrium, then the temperature in the box will follow the
Tolman law $T \propto 1/\chi$. Using $d\rho = Tds$ (and, hence,
$\partial \rho/\partial l = T \partial s/\partial l$), we see that
eq.\ (\ref{lower}) is equivalent to requiring that the entropy of the
box remain constant as it is lowered.}
\begin{equation}
0 = \frac{d E_{\infty}}{dl} - w = \int \frac{\partial \rho(l+y)}
{\partial l} \chi(l+y) dV.
\label{lower}
\end{equation}
Thus the work done by the weight of the box on the agent lowering it is
\begin{equation}
W_{g}(l) = - \int_{\infty}^{l} w(l') dl' = E_{i} - \int
\rho(l,y) \chi(l+y) dV
\end{equation}
where $E_{i}$ is the initial energy of the box.

Meanwhile, the thermal radiation exerts a buoyancy force on the box
equal to
\begin{equation}
f_{b}(l) = A [ (P \chi)_{l-b/2} - 
(P \chi)_{l+b/2}]
\end{equation}
where $P$ is the radiation pressure of the unconstrained thermal matter. 
The work done by the buoyancy force on the agent at infinity is then
\begin{equation}
W_{b}(l) = - \int_{\infty}^{l} f_{b}(l') dl' = - \int P(l,y) \chi(l+y) dV.
\end{equation}

If the contents of the box are dropped into the black hole from
position $l$, the increase in black hole entropy will be
\begin{equation}
\Delta S_{\rm{bh}} = \frac{1}{T_{\rm{bh}}} (E_{i} - W_{g} - W_{b}) =
\frac{1}{T_{\rm{bh}}} \int [\rho(l,y) + P(l,y)] \chi(l+y) dV.
\label{dsbh} 
\end{equation}
Using eq.\ (\ref{gfe}) together with $T = T_{\rm{bh}}/\chi$, we obtain
\begin{equation}
\Delta S_{\rm{bh}} = \frac{1}{T_{\rm{bh}}} \int
[\rho(l,y) - e(l,y)] \chi(l+y) dV + S_{\rm{th}}
\label{DS}
\end{equation}
where $S_{\rm{th}}$ is the entropy of the thermal radiation displaced
by the box. Equation~(\ref{DS}) is equivalent to eq.\ (20) of
\cite{bek94} and it corresponds directly to eq.\ (A12) of \cite{uw83} for
the process considered in that reference.

It also follows from eq.\ (\ref{gfe}) together with $T =
T_{\rm{bh}}/\chi$ that
\begin{equation} 
d(P \chi) = -e~d\chi .
\label{dif} 
\end{equation}
Minimizing $\Delta S_{\rm{bh}}$ with respect to $l$, and
using (\ref{dif}), we obtain
\begin{equation}
\int [\rho(l_{0},y) - e(l_{0},y)] \frac{\partial \chi(l+y)}
{\partial l} = 0 .
\label{fp} 
\end{equation}
Thus, the entropy increase of the black hole is minimal when the
contents are dropped in from the ``floating point'', i.e. when the
weight of the box is equal to the weight of the displaced thermal
radiation. Equation (\ref{fp}) is identical to eq.\ (14) of
\cite{bek94} and eq.\ (A13) of \cite{uw83}.

To proceed further, we first consider an idealized situation in which
we imagine that the box is filled with unconstrained thermal
matter.\footnote{It should be emphasized that we are not assuming here
that it is physically realistic to actually have a box filled with
unconstrained thermal matter. The consideration of such a box is done
here purely for mathematical purposes, to compare the generalized
entropy change that would occur in this idealized process to that
which occurs in the actual process (see below).} Let $T_0$ denote the
temperature of the matter in the box at the start of the process.
Then, when lowered to position $l$, the matter in the box will have a
temperature distribution $T = T_{\infty}(l)/\chi$, where
$T_{\infty}(l)$ is determined by $T_0$ and
eq.\ (\ref{lower}). According to our analysis above, the optimal place
(in the sense of minimizing $\Delta S_{\rm{bh}}$) to drop such a box
is at its ``floating point'', which is easily seen to be the position,
$l_0$, at which $T_{\infty}(l_0)$ = $T_{\rm bh}$, since at this
position we have $\rho = e$. By eq.\ (\ref{DS}), when the box
is dropped into the black hole from its floating point, $l = l_0$, we
have
\begin{equation}
\Delta S_{\rm{bh}} = S_{\rm{th}} = S
\label{imaginary}
\end{equation}
and there is no change in the generalized entropy. Consequently, if
the box is dropped from {\it any} position, $l$, we have
\begin{equation}
\Delta S' \geq 0
\end{equation}
and the GSL holds in this idealized process.

Now consider the actual process in which the box is filled with some
(arbitrary) distribution of matter, is lowered to an arbitrary
position $l$ (not necessarily the floating point of the box) and then
is dropped into the black hole. Let us compare the change in
generalized entropy in this process with the change in generalized
entropy that would occur in the above idealized process where we
choose $T_0$ so that at position $l$ the energies as measured at
infinity, $E_\infty$, of the two boxes agree. Then, it follows
immediately from eqs.\ (\ref{Einfty}) and (\ref{DS}) that the change in
black hole entropy, $\Delta S_{\rm{bh}}$, is the same for both
processes. However, since at position $l$ the boxes have the same
energy at infinity, the entropy, $S$, contained in the box in the
actual process cannot be larger than the entropy contained in the box
in the idealized process. Consequently, the change in generalized
entropy in the actual process cannot be smaller than the change in
generalized entropy in the idealized process, which was shown above to
be non-negative. This proves that the GSL cannot be violated in the
actual process.

\section{Bekenstein's analysis}

In \cite{bek94} Bekenstein purports to show that for thick boxes whose
``height'', $b$, is not small compared with $A^{1/2}$ (where, as above,
$A$ denotes the horizontal cross-sectional area of the box), the contents of 
the box must satisfy the entropy bound (\ref{S/E}) if the GSL is to
hold. We now briefly review Bekenstein's assumptions and conclusions, and
then reconcile them with the results of the previous section.

In his analysis, Bekenstein assumes that unconfined thermal matter can
be modelled as an $N$-species mixture of noninteracting massless
particles, so that
\begin{equation}
P = \frac{e}{3} = \frac{N \pi^2 T^4}{45}
\label{radiation}
\end{equation}
Bekenstein then makes the approximation\footnote{Equation
(\ref{xproptol}) is a good approximation sufficiently near the black
hole. Bekenstein's justification for this approximation is somewhat
circular in nature, but eq.\ (\ref{xproptol}) is not the source of any
difficulties in Bekenstein's analysis.} that
\begin{equation}
\chi(l) \approx \kappa l 
\label{xproptol} 
\end{equation}
where $\kappa$ denotes the surface gravity of the black hole. Using
this approximation, Bekenstein finds that the exact floating point
condition (\ref{fp}) reduces to
\begin{equation}
\frac{(l_{0}^{2} - b^2/4)^{3}}{3 l_{0}^{2} b^{4} + b^{6}/4} = 
\frac{N A}{720 \pi^2 E(l_{0}) b^{3}}
\label{float}
\end{equation}
where $E(l_{0}) = \int \rho dV$ is the locally-measured energy of the 
box at the floating point.

Bekenstein then argues that at the floating point, the quantity
\begin{equation}
\eta^3 \equiv \frac{N A}{720 \pi^2 E(l_{0}) b^{3}}
\label{eta}
\end{equation}
must satify $\eta \ll 1$. In making this argument, Bekenstein makes two
additional assumptions: (1) that $b \gg 1/E$ and (2) that $N$ is of
order unity. (It is easy to see that these assumptions together with
$A \lesssim b^{2}$ imply $\eta \ll 1$.) However, these assumptions are
not innocuous ones since, in conjunction with eq.\ (\ref{radiation})
they would imply that entropy bound (\ref{S/E}) is already satisfied
by a wide margin for a box in Minkowski spacetime. Namely, since the
box's contents must have lower entropy than unconstrained thermal
matter at the same energy and volume, we have for the model of
unconstrained thermal matter assumed by Bekenstein,
\begin{equation}
\frac{S}{E} \leq \left(\frac{S}{E}\right)_{\rm{th}} \sim
\frac{1}{T}
\end{equation}
Hence, given that $b \gg 1/E$, $N \sim 1$, and $A \lesssim b^{2}$, we
have
\begin{equation}
E \sim A b T^{4} \gg \frac{1}{b},
\end{equation}
from which it follows that
\begin{equation}
\frac{S}{E} \ll (A b^{2})^{1/4} \lesssim b = 2R.
\end{equation}

Nevertheless, Bekenstein's arguments correctly show
that---irrespective of the above two additional assumptions---if the
floating point of the box is very close to the horizon (in which case,
by eqs.\ (\ref{float}) and (\ref{eta}), we have $\eta \ll 1$), then
buoyancy effects are negligible, and the bound (\ref{S/E}) is needed
for the validity of the GSL. However, we now show that if
unconstrained thermal matter is described by (\ref{radiation}), then
any box that floats very close to the horizon must automatically satisfy
(\ref{S/E}).
Once again, we use the fact that
unconstrained thermal matter maximizes entropy at a fixed volume and
energy at infinity,
\begin{equation}
S(E_{\infty},l_{0}) \leq S_{\rm{th}}(E_{\infty},l_{0}) .
\label{entropy}
\end{equation}
The unconstrained thermal matter is described by eq.\ (\ref{radiation}) 
with $T = T_{\infty}(l_{0})/\chi$, where $T_{\infty}(l_{0})$ is determined by 
imposing $\int e \chi dV = E_{\infty}$. Evaluating this integral using the
approximation (\ref{xproptol}), we find
\begin{equation}
[T_{\infty}(l_{0})]^{4} = \frac{15 (l_{0}^{2} - b^{2}/4)^{2} \kappa^{3} 
E_{\infty}}{N \pi^{2} A b l_{0}}
\label{t0}
\end{equation}
The entropy density of the thermal radiation is $s = 4e/3T$, so
\begin{equation}
S_{\rm{th}}(E_{\infty}) = \frac{4 E_{\infty}}{3 T_{\infty}(l_{0})}.
\label{srad}
\end{equation}
It is convenient to express $E_{\infty}$ in terms of the position,
$l_{\rm{cm}}$, of the center of mass of the box. Again applying the 
$\chi \propto l$ approximation, we obtain the simple relation
\begin{equation}
l_{\rm{cm}} \equiv \frac{\int (l+y)~\rho~dV}{E(l_{0})} = 
\frac{E_{\infty}}{\kappa E(l_{0})}.
\label{lcm}
\end{equation}
By eqs.\ (\ref{entropy}), (\ref{srad}) and (\ref{lcm}), we have
\begin{equation}
\frac{S}{E} \leq \frac{8 \pi}{3} \left(
\frac{T_{\rm{bh}}}{T_{\infty}(l_{0})} \right) l_{\rm{cm}},
\end{equation}
and from eq.\ (\ref{t0}) and the definition, eq.\ (\ref{eta}), of $\eta$, 
we find
\begin{equation}
\left( \frac{T_{\rm{bh}}}{T_{\infty}(l_{0})} \right)^{4} =
\frac{3 \eta^{3} b^{4} l_{0}}{(l_{0}^{2} - b^{2}/4)^{2} l_{\rm{cm}}}.
\end{equation}
Now, assuming $\eta \ll 1$, the floating point condition (\ref{float})
yields $l_{0}^2 \approx (1/4 + \eta) b^{2}$. Consequently,
\begin{equation}
\frac{T_{\rm{bh}}}{T_{\infty}(l_{0})} \approx \left( 3 \eta \frac{l_{0}}
{l_{\rm{cm}}} \right)^{1/4},
\end{equation}
and, finally, to leading order in $\eta$,
\begin{equation}
\frac{S} {E} \leq \frac{8 \pi}{3} 
(3 \eta l_{\rm{cm}}^{3} l_{0})^{1/4} \leq \frac{8 \pi}{3} b (3 \eta)^{1/4} 
\ll b = 2R.
\end{equation}
Thus, we see that if the box floats very
near the horizon, it follows that the entropy bound (\ref{S/E}) is
already satisfied by a wide margin.
Consequently, the bound (\ref{S/E}) does not have to be postulated as
an additional requirement.

This research was supported in part by NSF grant PHY 95-14726 and ONR
grant N00014-96-1-0127 to the University of Chicago.

\end{document}